\documentclass{PoS}

\title{An improved model of color confinement}

\ShortTitle{Color confinement}

\author{\speaker{Daniel Zwanziger}
\\
        Physics Department, New York University, New York, NY 10003, USA\\
        E-mail: \email{dz2@nyu.edu}}

\abstract{	We consider the free energy $W[J] = W_k(H)$ of QCD coupled to an external source $J_\mu^b(x) = H_\mu^b \cos(k \cdot x)$, where $H_\mu^b$ is, by analogy with spin models, an external ``magnetic'' field with a color index that is modulated by a plane wave.  We report an optimal bound on $W_k(H)$ and an exact asymptotic expression for $W_k(H)$ at large $H$.  They imply confinement of color in the sense that the free energy per unit volume $W_k(H)/V$ and the average magnetization $m(k, H) ={1 \over V} {\p W_k(H) \over \p H}$ vanish in the limit of constant external field $k \to 0$.  Recent lattice data indicate a gluon propagator $D(k)$ which is non-zero, $D(0) \neq 0$, at $k = 0$.  This would imply a non-analyticity in $W_k(H)$ at $k = 0$.  We present a model that is consistent with the new results and exhibits (non)-analytic behavior.  Direct numerical tests of the bounds are proposed.}

\FullConference{The many faces of QCD\\
		November 2-5, 2010\\
		Gent Belgium}

\newcommand{\beqa}{\begin{eqnarray}}
\newcommand{\eeqa}{\end{eqnarray}}
\newcommand{\beq}{\begin{equation}}
\newcommand{\eeq}{\end{equation}}

\newcommand{\p}{\partial}

\begin{document}

\section{Introduction}

	Recent numerical studies on large lattices of the gluon propagator $D(k)$ in Landau gauge in 3 and 4 Euclidean dimensions, reviewed recently in \cite{Cucchieri:2010.03}, yield finite values for $D(0) \neq 0$ \cite{Cucchieri:2008.12} - \cite{Cucchieri:2007b}, in apparent disagreement with the theoretical expectation that $D(0) = 0$, originally obtained by Gribov~\cite{Gribov:1977wm}, and argued in~\cite{Zwanziger:1991}.  The argument~\cite{Zwanziger:1991} which leads to $D(0) = 0$, relies on the hypothesis that the free energy $W(J)$ in the presence of sources $J$ is analytic in $J$ at low momentum $k$.  That hypothesis should perhaps be dropped in view of the apparent disagreement with the lattice data.  This is of interest because a non-analyticity in the free energy is characteristic of a change of phase.  

	The free energy $W(J)$ enters the picture because it is the generating functional of the connected gluon correlators.  In particular the gluon propagator is a second derivative of $W(J)$ at $J = 0$,
\beq
D_{\mu \nu}^{ab}(x, y) = {\delta^2 W(J) \over \delta J_\mu^a(x) \delta J_\nu^b(y)}\Big|_{J=0}.
\eeq
The free energy $W(J)$ in the presence of sources $J$ is given by
\beqa
\exp W(J) & = & \langle \exp(J, A) \rangle
\nonumber \\
& = & \int_\Omega dA \ \rho(A) \exp(J, A),
\eeqa
where $\mu, \nu$ are Lorentz indices, and $a, b$ are color indices, and
\beq
(J, A) = \int d^Dx \ J_\mu^b(x) A_\mu^b(x).
\eeq
The integral over $A$ is effected in Landau gauge $\p_\mu A_\mu = 0$, and the domain of integration is restricted to the Gribov region $\Omega$, a region in $A$-space where the Faddeev-Popov operator is non-negative, $M(A) \equiv - \p_\mu D_\mu(A) \geq 0$.  We use continuum notation and results, but we have in mind the limit of lattice QCD in the scaling region, that is gauge-fixed to the Landau (or Coulomb) gauge by a numerical algorithm that minimizes the Hilbert norm squared $||A||^2$, and thereby fixes the gauge to the interior of the Gribov region.  The vector potential, given by $A(x) = g A^{\rm pert}(x)$, is unrenormalized, and has engineering dimension in mass units $[A(x)] = 1$ in all Euclidean dimension $D$, while $[H] = D-1$.  (Our results also hold in the Coulomb gauge at fixed time, in which case $D$ is the number of space dimensions.)   The density $\rho(A)$ is a positive, normalized probability distribution with support in the Gribov region $\Omega$.  Because there are Gribov copies inside $\Omega$, $\rho(A)$ is not unique and, in general, depends on the minimization algorithm.

	We consider a source that has the particular form
\beq
J_\mu^b(x) = H_\mu^b \cos(k x_1),
\eeq
where we have aligned the 1-axis along $k$, so the free energy
\beq
\label{freeenergy}
\exp W_k(H) =
\langle \ \exp[ \int d^Dx \ H_\mu^b \cos(k x_1) A_\mu^b(x) ] \ \rangle,
\eeq
depends only on the parameters $k$ and $H_\mu^b$.  This is sufficient to generate the gluon propagator for momentum $k$,
\beq
\label{Wtoprop}
D_{ij}^{ab}(k) = 2 \ {\p^2 w_k(0) \over \p H_i^a \p H_j^b },
\eeq
where
\beq
w_{k}(H) \equiv {W_k(H) \over V}
\eeq
is the free energy per unit Euclidean volume.  Because $A_\mu(x)$ is transverse, only the transverse part of $H$ is operative, and we impose $k_\mu H_\mu^b = 0$, which yields $H_1^a = 0$, and we write $H_i^a$, where $i = 2,... \ D$.  By analogy with spin models, $H_i^b$ may be interpreted as the strength of an external ``magnetic'' field, with a color index $b$, which is modulated by a plane wave $\cos(k x_1)$.  (This external magnetic field $H_i^b$, with color index $b$, should not be confused with the Yang-Mills color-magnetic field $F_{ij}^b$.)

	A rigorous bound for $W_k(H)$ on a finite lattice was given in \cite{Zwanziger:1991} which holds for {\it any} (numerical) gauge fixing with support inside the Gribov region $\Omega$.  One can easily show that  in the limit of large lattice volume $V$, and in the continuum limit, this implies the Lorentz-invariant continuum bound in $D$ Euclidean dimensions,
\beq
\label{LIcontbound}
w_{k}(H) \leq (2 D k^2)^{1/2} |H|,
\eeq
where $|H|^2 = \sum_{\mu, b} (H_\mu^b)^2$.  A model satisfying the bound (\ref{LIcontbound}) was recently exhibited in \cite{Zwanziger:2010b}.  

	More recently, a stricter bound for $w_k(H)$ at finite $H$ was obtained~\cite{Zwanziger:2011a}, that also holds for {\it any} (numerical) gauge fixing  with support inside the Gribov region $\Omega$, 
\beq
\label{newbound}
w_k(H) \leq 2^{-1/2} k \ {\rm tr}[ ( H^a H^a )^{1/2}].
\eeq
Here $H^a H^a$ is the matrix with elements $H_i^a H_j^a$.  It has positive eigenvalues, and the positive square root is understood.  A proof may be found in the Appendix that this bound is stricter than the old bound (\ref{LIcontbound}).  This bound is in fact optimal for a probability distribution $\rho(A)$ of which it is known only that its support lies inside the Gribov region.
	
	The same expression also provides the asymptotic form of $w_k(H)$ at large $H$, and infinite Euclidean volume $V$~\cite{Zwanziger:2011a} for any numerical gauge fixing with probability density $\rho(A)$ with support that reaches all boundary points of $\Omega$, (but which may vanish on the boundary, $\rho(A) = 0$ for $A \in \p\Omega$)
\beq
\label{Was}
w_{k, {\rm as}}(H) = 2^{-1/2} k \ {\rm tr}[ ( H^a H^a )^{1/2}].
\eeq

	Either bound yields in the zero-momentum limit
\beq
\label{0momentum} 
w_{0}(H) = \lim_{k \to 0} w_{k}(H) = 0.
\eeq
As discussed in \cite{Zwanziger:1991}, this states that the system does not respond to a constant external color-magnetic field {\em no matter how strong}.  It is a consequence of the proximity of the Gribov horizon in infrared directions.  We shall return in the concluding section to the physical implications of this result for confinement of color.

	If $w_k(H)$ were analytic in $H$ in the limit $k \to 0$, eq.~(\ref{0momentum}) would imply that all derivatives of the generating function $w_0(H)$ vanish, including in particular the gluon propagator (\ref{Wtoprop}) at $k = 0$, $D(0) = 0$.  However, as noted above, this disagrees with recent lattice data which indicate a finite value, $D(0) \neq 0$, in Euclidean dimensions 3 and 4.  If this is true, then $w_k(H)$ must become non-analytic in $H$ in the limit $k \to 0$.  In order to get some insight about this, we examine the behavior of an improved model that has the exact asymptotic behavior (\ref{Was}).

\section{Improved model}

	The model is defined by the expression for the free energy
\beq
\label{model}
w_{k,{\rm mod}}(H) = g(k) \Big\{ {\rm tr} \Big[ \Big(I + {k^2 H^a H^a \over 2 g^2(k)} \Big)^{1/2} - I \Big]
 - {\rm tr} \ln  \Big[ 2^{-1} \Big(I + {k^2 H^a H^a \over 2 g^2(k)} \Big)^{1/2} + 2^{-1} I  \Big] \Big\},
\eeq
where $g(k) \geq 0$ is an as yet undetermined function, and $H^a H^a$ is the matrix with elements $H_i^a H_j^a$, for $i, j = 2,... \ D$.   This model possesses the following desirable features \cite{Zwanziger:2011a}: (i) It satisfies $w_{k,{\rm mod}}(0) = 0$, which is correct at $H = 0$ for a normalized probability distribution $\int dA \ \rho(A) = 1$.  (ii) It has the asymptotic limit  
\beq
w_{k, {\rm as}}(H) = \lim_{\mu \to \infty} {w_{k, {\rm mod}}(\mu H) \over \mu}  = 2^{-1/2} k \ {\rm tr} [ ( H^a H^a )^{1/2} ] ,
\eeq
that is correct at large $H$ for any numerical gauge fixing that is strictly positive in the interior of the Gribov region $\Omega$. (iii) It satisfies the optimal bound
\beq
w_{k,{\rm mod}}(H) \leq 2^{-1/2} k \ {\rm tr} [ ( H^a H^a )^{1/2} ],
\eeq
which implies that the generating function vanishes at $k = 0$, $w_{0, {\rm mod}}(H) = 0$.  (iv) The matrix of second derivatives is positive in the sense that
\beq
v_i^a{\p^2 w_{k, {\rm mod}}(H) \over \p H_i^a \p H_j^b } v_j^b \geq 0
\eeq
holds for all $v_i^a$ and $H_i^a$, as required for this matrix to be a covariance.

	Because of the property $w_{0, {\rm mod}}(H) = 0 $, there must be some non-analyticity if, as indicated by numerical calculations, the gluon propagator $D(k)$ at $k = 0$ is positive $D(0) > 0$.  It is instructive to see what kind of analyticity this would be in our model.  Let $\lambda(H) > 0$ be the largest eigenvalue of the matrix $H_i^a H_j^a$.  Inspection of (\ref{model}) shows that $w_{k, {\rm mod}}(H)$ is analytic in $H$ inside a radius of convergence
\beq
\lambda(H) = {2 g^2(k) \over k^2}.
\eeq 
Moreover from (\ref{model}) we have at small $H$,
\beq
w_{k, {\rm mod}}(H) = { k^2 \over 8 g(k) } H_i^a H_i^a + g(k) O[ k^4 H^4 / g^4(k) ]. 
\eeq	 
For the gluon propagator $D(k) \sim {\p^2 w_{k, {\rm mod}}(0) \over \p H_i^a \p H_j^b} \sim k^2/ g(k)$ to be finite at $k = 0$, as suggested by the lattice data, we must have $g(k) = {\rm const} \ k^2$ near $k = 0$.  In this case the coefficient of the $H^4$ term is of order $1/k^2$, which diverges as $k \to 0$, as do all higher order coefficients.  Moreover the radius of convergence of the series expansion of $w_{k, {\rm mod}}(H)$ is  $\lambda(H) = O(k^2)$, which vanishes like $k^2$.  

	Suppose that $g(k)$ has a power law behavior $g(k) \sim k^\nu$ at $k = 0$.  Then the radius of convergence behaves like $\lambda(H) \sim k^{2\nu -2}$, which approaches 0 with $k$ for $\nu >1$.  The gluon propagator behaves like $D(k) \sim k^{2 - \nu}$, and $w_{k, {\rm mod}}(H)$ is non-analytic in $H$ at $k = 0$ when the propagator has a power law $D(k) \sim k^p$ with $p < 1$.  Gribov's original calculation gave $D(k) \sim k^2/m^4$ which corresponds to $g(k) = O(m^4)$, and $w_{k, {\rm mod}}(H)$ is analytic in $H$ at $k = 0$, with a radius of convergence $\lambda(H) = O(k^{-2}) \to \infty$ for $k \to 0$.

\section{Conclusion}

		By analogy with spin models, we define, for each momentum $k$, the analog of the bulk magnetization in the presence of the external ``magnetic'' field $H_i^a$,
\beq
M_i^a(k, H) =
 {\p W_k(H) \over \p H_i^a},
\eeq
which describes the reaction of the spin system to the external color-magnetic field.  Its physical meaning in gauge theory is apparent from (\ref{freeenergy}) which yields,
\beqa
M_\mu^b(k, H) 
& = & \langle \int d^Dx \ \cos(kx_1) A_i^b(x) \rangle_H
\nonumber  \\
 & = & (1/2) \langle \ a_i^b(k) + a_i^b(-k) \ \rangle_H.
\eeqa
Thus the ``bulk magnetization'' is in fact the $k$-th fourier component of the gauge field in the presence of the external magnetic  field.  We also define the magnetization per unit (Euclidean) volume
\beq
m_i^a(k, H) = {M_i^a(k, H) \over V},
\eeq
given by
\beq
m_i^a(k, H) =
 {\p w_k(H) \over \p H_i^a}.
\eeq

	The asymptotic free energy (\ref{Was}) determines the asymptotic magnetization per unit volume at large $H$,
\beqa
m_{i, {\rm as}}^a(k, H) & = &  {\p w_{k, {\rm as}}(H) \over \p H_i^a}
\nonumber  \\
& = & 2^{-1/2} k [(H^bH^b)^{-1/2}]_{ij} H_j^a.
\eeqa
Its magnitude is given by $(m_{i, {\rm as}}^am_{i, {\rm as}}^a)(k, H) = k^2/2$, and we obtain the simple formula
\beq
\lim_{H \to \infty} (m_i^am_i^a)(k, H) = k^2/2,
\eeq
which holds for any numerical gauge fixing with support extending up to the boundary of the Gribov region $\Omega$.
	
	We arrive at the remarkable conclusion that in the limit of constant external magnetic field, $k \to 0$, the color magnetization per unit volume vanishes, {\it no matter how strong the external magnetic field},
\beq
\lim_{k \to 0} \ \lim_{H \to \infty} m_i^a(k, H) = 0.
\eeq
Thus the system does not respond to a constant external color-magnetic field.  In this precise sense the color degree of freedom $m_i^b(k, H) = {1 \over 2V } \langle a_i^b(k) + a_i^b(-k) \rangle_H$ is absent at $k = 0$.  This conclusion holds whether or not the free energy $w_k(H)$ is analytic in $H$ in the limit $k \to 0$.  Lattice data would indicate that it is not analytic.  Besides reporting this result, we have presented a model, defined in (\ref{model}), which saturates the asymptotic limit (\ref{Was}), and exhibits confinement of color.  As we have seen, $W_{k, {\rm mod}}(H)$ may be either analytic in $H$, or not, at $k = 0$, depending on the behavior of $g(k)$ at $k = 0$, but in either case, the conclusion stands, that the constant color degree of freedom of the gauge field is confined.

	Equations (\ref{newbound}) and (\ref{Was}) may be checked numerically, at least in principle, by using the formula $\exp W_k(H) = \langle \exp[ \int d^Dx \  H_i^b \cos(kx_1) A_i^b(x)] \rangle$ to make a numerical determination of the generating function itself.  For large values of $H$ this may fluctuate too wildly.  Alternatively one may measure the magnetization from the formula $M_\mu^b(k, H) 
= \langle \int d^Dx \ \cos(kx_1) A_i^b(x) \rangle_H$, where the source term $H_i^b\cos(kx_1) A_i^b(x)$ is included in the action that one simulates.  This requires simulating the theory fixed in the Landau gauge instead of generating an ensemble from the gauge-invariant Wilson action then gauge fixing.  It may be convenient to do this by numerical simulation of stochastic quantization \cite{Nakamura:1993} because that avoids calculating the Faddeev-Popov determinant explicitly.

\bigskip
	  
{\bf Acknowledgements}\\
The author recalls with pleasure stimulating conversations with Attilio Cucchieri and Axel Maas that took place at the conference, The Many Faces of QCD, November 2-5, 2010, Gent, Belgium.  I am grateful to the conference organizers for the valuable opportunity which this conference provided.

\appendix

\section{The new bound is stricter than the old}

		To prove that the new bound is stricter we write the old bound (\ref{LIcontbound}), for the case that $k$ is aligned along the 1-axis, as $w_k(H) \leq B_1$, where
\beq
B_1 \equiv (2D)^{1/2}k [ {\rm tr} (H^a H^a) ]^{1/2}.
\eeq		
We now diagonalize the matrix $H_i^a H_j^a$ by a rotation in the $(D-1)$-dimensional space, with eigenvalues $\lambda_i$, so
\beq
B_1 = (2D)^{1/2}k \Big( \sum_{i = 2}^D \lambda_i \Big)^{1/2}.
\eeq   
The new bound (\ref{newbound}) reads $w_k(H) \leq B_2$ where
\beq
B_2 = 2^{-1/2} k \ \sum_{i = 2}^D \lambda_i^{1/2},
\eeq
and we must show $B_2 < B_1$.  (Note that the operations of square root and trace are interchanged.)  We have
\beq
B_1^2 = 2 D k^2  \sum_{i = 2}^D \lambda_i,
\eeq
and
\beqa
B_2^2 & = & 2^{-1} k^2 \ \sum_{i, j = 2}^D \lambda_i^{1/2} \lambda_j^{1/2}
\nonumber  \\
& = & 2^{-1} k^2 \ \Big( \sum_{i = 2}^D \lambda_i +  \sum_{i \neq j = 2}^D \lambda_i^{1/2} \lambda_j^{1/2} \Big)
\nonumber  \\
& \leq & 2^{-1} k^2 \ \Big( \sum_{i = 2}^D \lambda_i + 2^{-1} \sum_{i \neq j = 2}^D (\lambda_i + \lambda_j) \Big)
\nonumber  \\
& = & 2^{-2} k^2 \ \sum_{i, j = 2}^D (\lambda_i + \lambda_j)
\nonumber  \\
& = & 2^{-1} k^2 (D -1) \ \sum_{i= 2}^D \lambda_i
\nonumber  \\
& = & (4D)^{-1} (D -1) \ B_1^2.
\eeqa
We thus obtain $B_2^2 \leq (4D)^{-1} (D -1) \ B_1^2 < B_1^2$, so the new bound is stricter, as asserted.

\end{document}